\documentclass[aoas,MSNbibl,nameyear,seceqn,dvips]{arximspdf}
\usepackage{dcolumn}
\usepackage{graphicx}

\doi{10.1214/13-AOAS658} 
\volume{7}
\issue{4}
\pubyear{2013}
\firstpage{1940}
\lastpage{1959}

\makeatletter

\newcolumntype{d}[1]{D{.}{.}{#1}}

\newcommand{\Mm}{\mathrm{Mm}}

\newtheorem{corollary}{Corollary}
\newtheorem{lemma}{Lemma}

\newproclaim{strategy}{Strategy}

\newcommand{\gammab}{\bolds\gamma}
\newcommand{\thetab}{\bolds\theta}
\newcommand{\Vb}{\mathbf{V}}
\newcommand{\Ab}{\mathbf{A}}
\newcommand{\Sb}{\mathbf{S}}
\newcommand{\Xb}{\mathbf{X}}
\newcommand{\Ib}{\mathbf{I}}
\newcommand{\Mb}{\mathbf{M}}
\newcommand{\yb}{\mathbf{y}}
\newcommand{\Gb}{\mathbf{G}}
\newcommand{\Eb}{\mathbf{E}}
\newcommand{\Lb}{\mathbf{L}}
\newcommand{\pb}{\mathbf{p}}
\newcommand{\hb}{\mathbf{h}}
\newcommand{\zerob}{\mathbf{0}}

\makeatother

\begin{document}
\begin{frontmatter}

\title{Maximin and maximin-efficient event-related fMRI designs under
a nonlinear model}
\runtitle{fMRI designs for nonlinear model}

\begin{aug}
\author[A]{\fnms{Ming-Hung} \snm{Kao}\corref{}\ead[label=e1]{mkao3@asu.edu}},
\author[B]{\fnms{Dibyen} \snm{Majumdar}\ead[label=e2]{dibyen@uic.edu}},\\
\author[C]{\fnms{Abhyuday} \snm{Mandal}\thanksref{t1}\ead[label=e3]{amandal@stat.uga.edu}}
\and
\author[C]{\fnms{John} \snm{Stufken}\thanksref{t2}\ead[label=e4]{jstufken@stat.uga.edu}}
\runauthor{Kao, Majumdar, Mandal and Stufken}
\affiliation{Arizona State University, University of
Illinois at Chicago
and University~of Georgia}
\address[A]{M.-H. Kao\\
School of Mathematical\\
\quad and Statistical Sciences\\
Arizona State University\\
Tempe, Arizona 85287\\
USA\\
\printead{e1}}
\address[B]{D. Majumdar\\
Department of Mathematics, Statistics\\
\quad and Computer Science\\
University of Illinois at Chicago\\
Chicago, Illinois 60607\\
USA\\
\printead{e2}} 
\address[C]{A. Mandal\\
J. Stufken\\
Department of Statistics\\
University of Georgia\\
Athens, Georgia 30602\\
USA\\
\printead{e3}\\
\hphantom{E-mail: }\printead*{e4}}
\end{aug}

\thankstext{t1}{Supported in part by NSF Grant DMS-09-05731 and NSA
Grant H98230-13-1-0251.}
\thankstext{t2}{Supported in part by NSF Grants DMS-07-06917 and DMS-10-07507.}

\received{\smonth{2} \syear{2013}}
\revised{\smonth{5} \syear{2013}}

%
\begin{abstract}
Previous studies on event-related functional magnetic resonance imaging
experimental designs are primarily based on linear models, in which a
known shape of the hemodynamic response function (HRF) is assumed.
However, the HRF shape is usually uncertain at the design stage. To
address this issue, we consider a nonlinear model to accommodate a wide
spectrum of feasible HRF shapes, and propose efficient approaches for
obtaining maximin and maximin-efficient designs. Our approaches involve
a reduction in the parameter space and a search algorithm that helps to
efficiently search over a restricted class of designs for good designs.
The obtained designs are compared with traditional designs widely used
in practice. We also demonstrate the usefulness of our approaches via a
motivating example.
\end{abstract}

%
\begin{keyword}
\kwd{A-optimality}
\kwd{cyclic permutation}
\kwd{design efficiency}
\kwd{genetic algorithms}
\kwd{hemodynamic response function}
\kwd{information matrix}
\end{keyword}

\end{frontmatter}

\section{Introduction}
Functional magnetic resonance imaging (fMRI) is a pioneering,
noninvasive brain mapping technology for studying brain functions
[\citet{Culham2006bk51,DEspositoetal1999ar51}]. It is
arguably one of the most important advances in neuroscience and has
many important clinical potentials such as early identification of
Alzheimer's disease, pre-neurosurgical planning, and post-neurosurgical
evaluations; see, \citet{Bookheimer2007ar51} and
\citet{WierengaBondi2007ar51}. This cutting-edge technology has
been applied in a wide variety of disciplines
[\citet{Lazar2008bk51,Lindquist2008ar51}].

In a typical fMRI experiment, a predetermined sequence of mental
stimuli (e.g., pictures or sounds)
is presented to a subject. While the subject is exposed to the stimuli,
an MR scanner repeatedly scans
the subject's brain to collect a blood oxygenated level dependent
(BOLD) time series from each brain voxel
(three-dimensional imaging unit). A study usually involves multiple
(e.g., $64 \times64 \times30$) voxels,
resulting in multiple time series. These time series reflect the MR
signal changes evoked by the underlying
brain activity and are analyzed to make statistical inference about the
inner workings of the brain.
A crucial first step for rendering a valid and precise inference is to
select a high quality experimental
design for the fMRI experiment.

Here, we focus on event-related (ER) fMRI designs with brief mental
stimuli. Such designs are very popular due to their flexibility [\citet
{Huettel2012ar51,JosephsHenson1999ar51}], and
are the primary focus of existing research on fMRI designs [e.g., \citet
{Mausetal2010ar51,Kaoetal2009ar51,Liu2004ar51,WagerNichols2003ar51}].
Current knowledge about
the performance of ER-fMRI designs is mainly based on general linear
models. While popular, the use of general linear models is criticized
by some researchers [\citet
{Lohetal2008ar51,WorsleyTaylor2006ar51,Handwerkeretal2004ar51}].
A major criticism is the assumption of a fixed, known shape
of the hemodynamic response function (HRF), a function of time
describing the noise-free MR signal
change evoked by one, single stimulus. This assumption is not always
valid. Studies showed that the HRF
shape may vary across brain voxels, and that a misspecified shape can
lead to incorrect conclusions.
To allow for uncertain HRF shapes, analysis methods such as the use of
nonlinear models have been
seen in the literature [e.g., \citet
{LindquistWager2007ar51,Handwerkeretal2004ar51,Miezinetal2000ar51}].
However, not much work has been done to address this important issue at
the design stage.

\citet{Kao2009bk51} investigated the performance of ER-fMRI designs
under a nonlinear model (Section~\ref{SecModel}) that can
accommodate a wide variety of feasible HRF shapes. With such a model,
the optimality criterion for evaluating the performance of designs
typically depends on unknown model parameters. \citet{Kao2009bk51}
assumed the availability of a prior distribution of
unknown parameters and put forward an approach for obtaining designs
optimizing a (pseudo-)Bayesian design criterion, which is the expected
value of the specific optimality criterion. \citet{Mausetal2012ar51}
considered a maximin-type approach
that focuses on the worst case scenario over a prespecified parameter
space containing possible values of
the model parameters. Specifically, they targeted designs that maximize
the worst
relative efficiency over the parameter space. Here, a relative
efficiency is the relative
value of the specific optimality criterion with respect to a locally
optimal design that
is optimal for a given parameter vector value. Following \citet
{Muller1995ar2071},
the obtained designs will be termed as maximin-efficient designs.

In contrast to maximin-efficient designs, maximin designs optimize the
worst value
of the optimality criterion. In other words, the maximin criterion focuses
directly on the worst performance of designs over the parameter space, and
the maximin-efficient criterion can be viewed as a ``weighted'' version
of maximin criterion.
The weights are determined by locally optimal designs or, more precisely,
the best possible value of the optimality criterion evaluated at each
parameter vector value.
Both criteria are considered in a wide variety of design problems [e.g.,
\citet
{BergerWong2009bk2021,Chenetal2008ar2071,HuangLin2006ar2071,Bergeretal2000ar2071,KingWong2000ar2071,Sitter1992ar2071,Silvey1980bk2021}].
Unfortunately, obtaining maximin-type designs optimizing these
criteria is very challenging. One typically needs to deal with an
optimization problem that is mathematically intractable and
computationally difficult, if not infeasible [\citet
{Chenetal2011ar2071,Detteetal2007ar2071}].
An efficient approach is thus crucially important.

In this paper, we propose approaches for obtaining
maximin and maximin-efficient designs for fMRI experiments to allow
uncertain HRF shapes.
We derive useful results and develop efficient strategies for obtaining
high-quality designs. Our strategies involve a reduced parameter space,
a restricted class
of ER-fMRI designs, and an efficient search algorithm for searching
over the restricted design class
for good designs. The usefulness of our approaches is demonstrated via
case studies
and a real example.

We note that \citet{Mausetal2012ar51} obtained $D$-optimal
maximin-efficient designs for one stimulus type. Here, we develop approaches
for obtaining both maximin and maximin-efficient designs, and apply our
methods to find
maximin-type designs under $A$-optimality for cases with one or more
stimulus types. $D$-optimal designs
help to control the volume of a confidence ellipsoid of the parameters.
By contrast, $A$-optimality
aims at maximizing the average estimation precision. While
$D$-optimality is not uncommon in fMRI,
the $A$-optimality criterion is widely accepted by researchers in the field
[see also \citet
{Mausetal2010ar51,Kaoetal2009ar51,Dale1999ar51,Fristonetal1999ar51}].
We also note that our proposed
methods can be applied to all optimality criteria that are invariant
under simultaneous permutation of
rows and columns of the information matrix. Both $A$- and
$D$-optimality criteria possess this
invariance property.

The remainder of the article is organized as follows. In Section~\ref{sec2} we
provide a brief introduction about ER-fMRI designs. We then introduce
our methods, including the underlying statistical model, optimality criteria,
and our proposed strategies for obtaining maximin and maximin-efficient designs.
Case studies and a real example are provided in Section~\ref{SecCase}. The paper
closes with
a conclusion in Section~\ref{sec4}.

\section{Background and methodology}\label{sec2}
\subsection{A nonlinear model} \label{SecModel}

An ER-fMRI design is a finite sequence of brief stimuli interlaced with
control to be presented to an experimental subject.
Each stimulus may last several milliseconds to a few seconds. Times
between consecutive stimulus onsets
are multiples of a prespecified time, called the inter stimulus
interval (ISI;
e.g., 4 s). The control (e.g., periods of fixation or rest) fills in
the time when no stimulus is being presented.
We may use a sequence of finite numbers, for example, $d = \{1 0 1 2 1 0
\cdots1\}$, to represent an ER-fMRI design. An integer $q$ $(\neq0)$
at the $k$th position indicates an onset of a
$q$th-type stimulus at time $(k-1)\mathrm{ISI}$. A~``$0$'' means no stimulus onset
at that time point.

%
\begin{figure}

\includegraphics{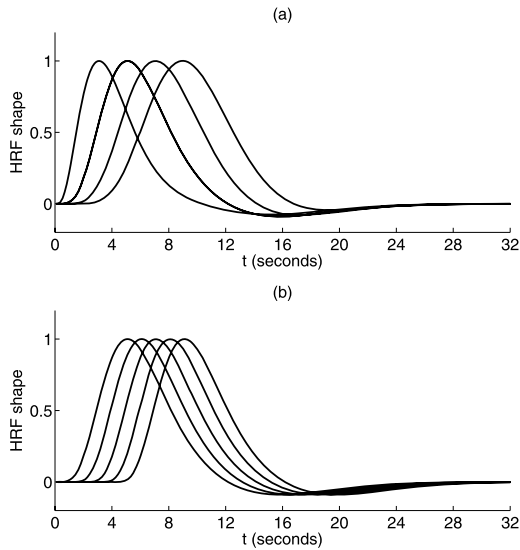}

\caption{The HRF shapes $g(t;\pb=(p_1,p_6))$ of (\protect\ref
{EqHRF}) with, from left to right, \textup{(a)} $p_6 = 0$ and $p_1 = 4$ to 10 in
steps of 2; and
\textup{(b)} $p_1 = 6$ and $p_6 = 0$ to 4 in steps of 1.} \label{FigHRF}
\end{figure}

At an activated brain voxel, each stimulus evokes a change in the MR
signal intensity.
The signal intensity takes about 25 to 30 seconds to rise and decay.
This change is typically described by an HRF having an assumed shape
with an unknown amplitude (maximal height);
see Figure~\ref{FigHRF} for some possible HRF shapes. When the next
stimulus occurs before the cessation of the current HRF,
the evoked HRFs accumulate. Along with nuisance signals and noise, the
accumulated HRF is acquired
by an MR scanner every $\mathrm{TR}$ (time-to-repetition; e.g., 2 s) to form the
BOLD time series.
Denoting the time series of a voxel by a $T$-by-$1$ vector $\yb$, we
consider the following
nonlinear model:
%
%
\begin{equation}\label{EqNLM}
\yb= \sum_{q = 1}^Q \Xb_{d, q} \hb(
\pb) \theta_q + \Sb\gammab+ \mathbf{e}.
\end{equation}
Here, $Q$ is the number of stimulus types. $\Xb_{d, q} \hb(\pb)
\theta_q$ represents
the accumulated HRF evoked by the $q$th-type stimuli of a design $d$. The
scalar $\theta_q$ is the unknown HRF amplitude. The vector $\hb(\pb)$,
indexed by an unknown parameter vector $\pb$, depicts the heights of
the HRF shape
after every $\Delta T$ seconds following a stimulus onset; $\Delta T$ is
the greatest value making both $(\mathrm{ISI}/\Delta T)$ and $(\mathrm{TR}/\Delta T)$ integers.
$\Xb_{d, q}$ is the 0--1 design matrix with 1 indicating the heights of
the HRF
that contribute to each BOLD measurement; a construction of $\Xb_{d,
q}$ can be found in
the Appendix of \citet{Kaoetal2012ar51}. The nuisance term $\Sb\gammab
$ allows a drift/trend
over time with an unknown parameter vector $\gammab$. The
correlated noise is represented by $\mathbf{e}$. For detecting
brain voxels activated by the stimuli, the focus is typically
on the amplitudes, $\thetab= (\theta_1,\ldots, \theta_Q)$, which
reflect the ``strengths'' of brain activation. A large $\theta_q$-value
signals a voxel that is highly activated
by the $q$th-type stimuli, $q = 1,\ldots, Q$.

With unknown $\pb$, model (\ref{EqNLM}) allows for an uncertain HRF
shape $\hb(\pb)$.
The vector $\hb(\pb)$ is determined by a continuous function $g(t;
\pb)$ with $t$ representing
time elapsed after a stimulus onset. There are many choices for $g(t;
\pb)$. Our selected $g(t; \pb)$
has the same form as the double-gamma function of SPM
(\url{http://www.fil.ion.ucl.ac.uk/spm/}), a popular computer software
package for
analyzing fMRI data:
%
%
\begin{equation}\label{EqHRF}
g(t; \pb) = \frac{g_0(t; \pb)}{\max_s g_0(s; \pb)},
\end{equation}
where
\begin{eqnarray*}
g_0(t; \pb) &=& f\biggl(t-p_6, \frac{p_1}{p_3},
p_3\biggr) - p_5 f\biggl(t-p_6,
\frac
{p_2}{p_4}, p_4\biggr);\\
f(x,\alpha, \beta) &=&
\frac{x^{\alpha- 1}
e^{-x/\beta}}{\Gamma(\alpha) \beta^\alpha};
\end{eqnarray*}
$\Gamma(\cdot)$ is the gamma function; and $f(x, \alpha, \beta)$ is
the probability density function of the gamma distribution,
$\operatorname{gamma}(\alpha, \beta)$. The double-gamma function of
SPM fixes $(p_1, p_2,\ldots, p_6) = (6, 16, 1, 1, 1/6, 0)$.
This function is completely known and is commonly used in the general
linear model approach for describing the HRF shape.
By contrast, we allow an uncertain HRF shape and follow \citet
{Wageretal2005ar51} to treat the two most influential HRF parameters,
namely, $p_1$, time-to-peak, and $p_6$, time-to-onset, as free
parameters, while keeping
the less sensitive parameters ($p_2, p_3, p_4, p_5$) fixed at $(16, 1,
1, 1/6)$. The HRF shapes
with selected $(p_1, p_6)$-values can be found in Figure~\ref{FigHRF}.
For brevity, we will omit the fixed
parameters $p_2, p_3, p_4$, and $p_5$ from $\pb$ and write $\pb=
(p_1, p_6)$, although $\pb$ should
really include six parameters. The $j$th element of the vector
$\hb(\pb)$ is then $g((j-1)\times(\Delta T); \pb)$. The length of
$\hb(\pb)$ is set to $1+\lfloor32/\Delta T \rfloor$ since
a typical HRF is nearly zero after 32 seconds; here, $\lfloor a \rfloor
$ is the integer part of $a$.

\subsection{Optimality criteria}\label{SecCriteria}
We aim at a good design for detecting activation (or studying $\thetab
$) with model (\ref{EqNLM}). The performance of a
design will be evaluated by $1/\operatorname{trace}(\operatorname{Cov}[\hat{\thetab}])$, the
reciprocal of the average variance of the
generalized least squares estimators $\hat{\thetab}$, that is,
$A$-optimality. Following a popular technique
[\citet{FedorovHackl1997bk2051,BoxLucas1959ar2051}],
we first linearize model (\ref{EqNLM}) and then use the linearized
model to approximate
$\operatorname{Cov}[\hat{\thetab}]$. The approximated covariance matrix
is proportional to $\Mb^{-1}(d; \thetab, \pb)$, where
\begin{eqnarray*}
\Mb(d; \thetab, \pb) &=& \Eb_d(\pb)' \bigl[
\Ib_T - w\bigl\{\Lb _d(\thetab, \pb)\bigr\} \bigr]
\Eb_d(\pb),
\\
\Eb_d(\pb) &=& \bigl[\Ib_T - w\{\Vb\Sb\} \bigr]\Vb
\Xb_d \bigl[\Ib _Q \otimes\hb(\pb) \bigr],
\\
\Lb_d(\thetab, \pb) &=& [\Lb_1, \Lb_6],\\
\Lb_i &=& \bigl[\Ib_T - w\{\Vb\Sb\} \bigr] \Vb
\Xb_d \biggl[\Ib_Q \otimes\frac{\partial\hb(\pb)}{\partial p_i} \biggr]
\thetab',\qquad i=1, 6,
\end{eqnarray*}
$\Ib_a$ is the $a$-by-$a$ identity matrix, $w\{\Ab\} =
\Ab(\Ab'\Ab)^{-}\Ab'$ is the orthogonal projection matrix onto the
column space of $\Ab$, $\Ab^{-}$ is a generalized inverse matrix
of~$\Ab$, $\Xb_d = [\Xb_{d, 1},\ldots, \Xb_{d, Q}]$, $\Vb$ is selected
so that $\Vb\mathbf{e}$ is white noise, $\otimes$ is the
Kronecker product, and the vector ($\partial\hb(\pb )/\partial p_i$) is
determined by the partial derivative of $g(t; \pb)$ with respect to
$p_i$, $i = 1, 6$.

We would like a design maximizing $\Phi_A(d; \thetab, \pb) \equiv
1/\operatorname{trace}(\Mb^{-1}(d; \thetab, \pb))$.
The answer will depend on the unknown $\thetab$ and $\pb$. This makes
such a nonlinear design
problem notoriously difficult. One way for tackling such a problem is by
obtaining a locally optimal design [\citet{Chernoff1953ar2051}]
that is optimal for a given $(\thetab, \pb)$-value. However, this
approach is unsatisfactory for fMRI.
This is because a good guess for the parameter vector value is almost
always unavailable. More importantly, the selected design should be
efficient for
the various parameter values (or HRF shapes) associated with all the
brain voxels of interest.
We thus resort to the maximin and maximin-efficient approaches.

The maximin approach seeks designs maximizing
%
%
\begin{equation}\label{EqMm}
\min_{(\thetab, \pb) \in\Theta\times\mathcal{P}} \Phi_A(d; \thetab, \pb),
\end{equation}
where $\Theta\times\mathcal{P}$ is a specified parameter space of
$(\thetab, \pb)$.
A maximin design thus maximizes the worst average precision in estimating
$\thetab$ by taking the uncertainty of both $\thetab$ and $\pb$ into
account. On the other hand,
the maximin-efficient criterion is
%
%
\begin{equation}\label{EqMmE}
\min_{(\thetab, \pb) \in\Theta\times\mathcal{P}} \operatorname{RE}(d; \thetab, \pb) = \min_{(\thetab, \pb) \in\Theta\times\mathcal{P}}
\frac{\Phi
_A(d; \thetab, \pb)}{\Phi_A(d^*_{\thetab, \pb}; \thetab, \pb)},
\end{equation}
where\vspace*{1pt} $d^*_{\thetab, \pb}$ is a locally optimal design maximizing
$\Phi_A$ for given $(\thetab, \pb)$.
To reflect that the HRF typically increases in 0--2 s after the stimulus
onset, reaches the peak in 5--8 s, and then falls back to baseline
[\citet{Lindquist2008ar51,Rosenetal1998ar51}], we set $\mathcal{P} =
\{(p_1, p_6) \mid p_1 \in[6,9], p_6 \in[0,2]\}$. This choice also
follows the fact that the mode of the gamma distribution $\operatorname{gamma}(\alpha,
1)$ is $(\alpha-1)$ for $\alpha> 1$. Other $\mathcal{P}$ can also be
considered. With no further information, we consider $\mathbb{R}^Q$ as
the parameter space of $\thetab$, which can be greatly reduced using
the results presented in the next subsection.

\subsection{Strategies to find maximin and maximin-efficient designs}
\label{Secmethod}
Obtaining maximin or maximin-efficient ER-fMRI designs is
computationally challenging.
Results in this section help to reduce the computational burden. We
first discuss results useful for the maximin approach. Some
of these results can also be applied to the maximin-efficient approach.
Additional
results that facilitate the maximin-efficient approach are then described.

\begin{lemma}\label{lemma1} $\Mb^{-1}(d; \zerob, \pb) \leq\Mb^{-1}(d; \thetab,
\pb)$ in L\"{o}wner ordering for any $\thetab$, $\pb$, and a design $d$
that ensures the existence of $\Mb^{-1}(d; \thetab, \pb)$.
\end{lemma}

\begin{lemma}\label{lemma2}
$\Mb(d; c \thetab, \pb) = \Mb(d; \thetab, \pb)$
for any scalar $c \neq0$.
\end{lemma}

The first lemma follows from Theorem 18.3.4 of \citet
{Harville1997bk31}, and allows us to leave out $\zerob$ from the
parameter space of $\thetab$ when obtaining maximin designs. We note
that the existence of $\Mb^{-1}(d; \zerob, \pb)$ is guaranteed
by the nonsingularity of $\Mb(d; \thetab, \pb)$. Lemma \ref{lemma2} is linked
to an observation made by \citet{BoseStufken2007ar41}. It
suggests that the $\Phi_A$-value depends on the direction of $\thetab
$, but not on its length. Thus,
when $Q = 1$, $\Phi_A(d; \theta_1, \pb) = \Phi_A(d; 1, \pb)$ for
any $\pb$ and $\theta_1 \neq0$.
The parameter space can then be reduced to $\{1\} \times\mathcal{P}$
from $\mathbb{R} \times\mathcal{P}$.
For $Q > 1$, we represent $\thetab$ using the hyper-spherical
coordinate system, and focus only on
the surface of the $Q$-dimensional unit hemisphere centered at the
origin. Specifically, for $Q = 2$, the
parameter space of $\thetab$ can be reduced to $\Theta= \{(\cos
\varphi_1, \sin\varphi_1) \mid\varphi_1 \in(-\pi/2, \pi/2]\}$.
For $Q=3$, $\Theta= \{(\cos\varphi_1, \sin\varphi_1 \cos\varphi
_2, \sin\varphi_1 \sin\varphi_2)
\mid\varphi_i \in(-\pi/2, \pi/2]\}$ can be used. For a larger $Q$,
we have $\Theta= \{(\theta_1,\ldots, \theta_Q)\}$, where
\begin{eqnarray*}
\theta_1 &=& \cos\varphi_1;\qquad \theta_q = \cos
\varphi_q \prod_{i=1}^{q-1} \sin
\varphi_i,\qquad q = 2,\ldots, Q-1;
\\
\theta_Q &=& \prod_{i=1}^{Q-1}
\sin\varphi_i; \varphi_1,\ldots, \varphi_{Q-1}
\in(-\pi/2, \pi/2].
\end{eqnarray*}
The two lemmas allow for a large reduction in the parameter space and
facilitate the search for maximin designs.
To further decrease the computational cost, we propose an efficient
strategy using the following result.

\begin{lemma}\label{lemma3}
Let $\mathcal{G} = \{\Gb_1,\ldots, \Gb_G\}$ be a set of
$Q \times Q$ permutation \mbox{matrices}. Suppose
$\Theta_0 \subset\Theta$ is such that $\Theta= \bigcup_{g = 0}^{G}
\Theta_g$, where
$\Theta_g = \{ \Gb_g \thetab\mid\thetab\in\Theta_0\}$ and $\Gb
_0 \equiv\Ib_Q$.
If $d_{\Mm, \Theta_0}$ is a maximin design for $\Theta_0 \times
\mathcal{P}$ and $
\min_{\Theta_0\times\mathcal{P}} \Phi_A(d_{\Mm, \Theta_0};\allowbreak \thetab, \pb) = \min_{\Theta_g
\times\mathcal{P}} \Phi_A(d_{\Mm, \Theta_0}; \thetab, \pb)$ for
any $g$, then $d_{\Mm, \Theta_0}$
is also a maximin design for $\Theta\times\mathcal{P}$.
\end{lemma}

A proof of Lemma \ref{lemma3} can be found in the \hyperref[app]{Appendix}. It is noteworthy that,
although we present Lemma \ref{lemma3} using $\Phi_A$,
this lemma can be applied to any optimality criterion that is invariant
under simultaneous permutation of rows and columns
of the information matrix $\Mb(d; \thetab, \pb)$. Many commonly used
optimality criteria, including $A$- and $D$-optimality, satisfy this
invariance property; see also \citet{Cheng1996bk2021}.
This lemma motivates the following strategy for obtaining maximin designs:

\begin{strategy}\label{strategy1}
(1) Identify a $\Theta_0$ and $\mathcal{G}$; and
(2) obtain a design $d_{\Mm, \Theta_0}$ maximizing $\min_{\Theta
_0\times\mathcal{P}} \Phi_A(d; \thetab, \pb)$, for which the ratio
%
%
\begin{equation}\label{Eqg0ratio}
\mathcal{R}_g = \frac{\min_{\Theta_g\times\mathcal{P}} \Phi
_A(d_{\Mm, \Theta_0}; \thetab, \pb)}{\min_{\Theta_0 \times\mathcal
{P}} \Phi_A(d_{\Mm, \Theta_0}; \thetab, \pb)}
\end{equation}
is 1 for any $g=1,\ldots, G$.
\end{strategy}

If such a $d_{\Mm, \Theta_0}$ exists, then
it is a maximin design for the entire parameter space. On the other hand,
if $\mathcal{R}_g < 1$ for some $g$, calculating the minimal $\mathcal
{R}_g$ still provides
a lower bound for the efficiency of $d_{\Mm, \Theta_0}$. More precisely,
%
%
\begin{eqnarray}\label{EqRg}
\min_{g\neq0} \mathcal{R}_g &\leq& \min
_{g\neq0} \frac{\min_{\Theta_g\times\mathcal{P}} \Phi_A(d_{\Mm, \Theta_0}; \thetab,
\pb)}{
\min_{\Theta_0 \times\mathcal{P}} \Phi_A(d_{\Mm}; \thetab, \pb)}
\nonumber\\
&\leq& \min_{g\neq0} \frac{\min_{\Theta_g\times\mathcal{P}} \Phi
_A(d_{\Mm, \Theta_0}; \thetab, \pb)}{\min_{\Theta\times\mathcal
{P}} \Phi_A(d_{\Mm}; \thetab, \pb)} \\
&=& \frac{\min_{\Theta\times\mathcal{P}} \Phi_A(d_{\Mm, \Theta_0};
\thetab, \pb)}{\min_{\Theta\times\mathcal{P}} \Phi_A(d_{\Mm};
\thetab, \pb)},
\nonumber
\end{eqnarray}
where $d_{\Mm}$ is a maximin design for $\Theta\times\mathcal{P}$. Note
that the equality in (\ref{EqRg}) follows from the fact that
$\min_{\Theta _0\times\mathcal{P}} \Phi_A(d_{\Mm, \Theta_0}; \thetab,
\pb) \geq \min_{\Theta_g\times\mathcal{P}} \Phi_A(d_{\Mm,
\Theta_0};\break
\thetab, \pb)$ for any $g$, which can be proved by using Lemmas
\ref{lemmaA.1} and \ref{lemmaA.2} in the \hyperref[app]{Appendix}. If
the minimal $\mathcal{R}_g$ is close to 1, $d_{\Mm, \Theta_0}$ will
perform well in terms of the maximin criterion~(\ref{EqMm}).

We now turn to results that help to obtain maximin-efficient designs.
To compute
the $\mathrm{RE}$-value in (\ref{EqMmE}), we need locally optimal designs for
all $(\thetab, \pb)$
in the parameter space. Obtaining these locally optimal designs is
computationally demanding (or infeasible),
especially when the parameter space is large. The following results
partly relieve this computational burden.

\begin{corollary}\label{coroll1}
A locally optimal design $d^*_{\thetab, \pb}$ for
$(\thetab, \pb)$ is also a locally optimum design for
$(c\thetab, \pb)$ for any $c \neq0$.
\end{corollary}

Corollary \ref{coroll1} follows from Lemma \ref{lemma2}. We also have the following corollary that
allows for a reduction in the parameter space when obtaining
maximin-efficient designs.

\begin{corollary}\label{coroll2}
$\operatorname{RE}(d; \thetab, \pb) = \operatorname{RE}(d; c\thetab, \pb)$ for
any design $d$ and $c \neq0$.
\end{corollary}

With this corollary, we may now reduce the parameter space to
$ \{ \{\zerob\} \cup\Theta \} \times\mathcal{P}$ with
$\Theta$ being the surface of the $Q$-dimensional unit hemisphere
centered at the origin.
Similarly to Lemma \ref{lemma3}, we make the following observation to help further
reduce the parameter space.

\begin{lemma}\label{lemma4}
A maximin-efficient design $d_{\mathrm{MmE}, \Theta_0}$ for
$ \{ \{\zerob\} \cup\Theta_0  \} \times\mathcal{P}$ is
also a maximin-efficient design for $\Theta\times\mathcal{P}$ if,
for any g,
\[
\min_{ \{ \{\zerob\} \cup\Theta_0  \} \times\mathcal
{P}} \operatorname{RE}(d_{\mathrm{MmE}, \Theta_0}; \thetab, \pb) = \min
_{ \{ \{\zerob\} \cup\Theta_g  \}
\times\mathcal{P}} \operatorname{RE}(d_{\mathrm{MmE}, \Theta_0}; \thetab, \pb).
\]
\end{lemma}

With Lemma \ref{lemma4}, a strategy similar to Strategy \ref{strategy1} can be considered
for obtaining maximin-efficient designs. Specifically, we may find
a maximin-efficient design for the reduced parameter space
$ \{ \{\zerob\} \cup\Theta_0  \} \times\mathcal{P}$,
and check if the minimal $\mathrm{RE}$-value of the obtained design is similar
across all $ \{ \{\zerob\} \cup\Theta_g  \} \times
\mathcal{P}$, $g = 0, 1,\ldots, G$.
Unfortunately, this strategy, which works well for the maximin
approach, may fail to provide
good maximin-efficient designs; see Section~\ref{SecCase}. A closer
look reveals that
the $\mathrm{RE}$-values and $\min$-$\mathrm{RE}$ can be greatly changed after a
permutation of the coordinates of $\thetab$.
This motivates us to consider another strategy suggested by Lemmas
\ref{lemma3}
and \ref{lemma4}. The idea is
to search for maximin-type designs over a restricted class of designs
for which
the $\Phi_A$-values are (nearly) invariant to permutations of the
elements of $\thetab$.
The restricted design class $\Xi_0$ that we consider is described below.

With $Q (> 1)$ stimulus types and a design length $L$, each design in
the restricted design class
$\Xi_0$ is formed by a ``short design'' of length $\lceil L/Q \rceil$;
$\lceil a \rceil$
is the smallest integer $\geq a$. The labels
of stimulus types of the initial short design are cyclically permuted to
generate additional $Q-1$ short designs. In particular, the label $q$
in the current short design is replaced by $q+1$ in the next short
design, \mbox{$q = 1,\ldots, Q-1$};
the label $Q$ becomes 1, and $0$'s are kept intact. A design of length
$L$ is then achieved
by concatenating the $Q$ short designs and leaving out the last $(Q
\lceil L/Q \rceil- L)$ elements.
The design class $\Xi_0$ has
also been considered by \citet{Kaoetal2009ar52}. Here, we are able to
show in the supplementary document [\citet{Kaoetal2013ar51}] that, with
a simplified model and two stimulus types
($Q = 2$), the $\Phi_A$-values of designs in $\Xi_0$ are quite
insensitive to permutations of the elements of $\thetab$.
Based on our empirical results, this observation tends to remain true
for more
realistic situations. We now describe our proposed second strategy.\vadjust{\goodbreak}

\begin{strategy}\label{strategy2}
(1) Identify a $\Theta_0$ and $\mathcal{G}$; and
(2) obtain a design $d_{\mathrm{MmE}, \Theta_0}$
that maximizes $\min_{\{ \{\zerob\} \cup\Theta_0 \} \times
\mathcal{P}} \operatorname{RE}(d; \thetab, \pb)$ in the subclass $\Xi_0$.
\end{strategy}

The results of our case studies indicate that good maximin-efficient
designs can be found by Strategy \ref{strategy2} with a greatly reduced computing
time. We note that, by considering the maximin criterion of (\ref
{EqMm}) in
Strategy \ref{strategy2}, maximin designs can be obtained over the subclass $\Xi_0$.
In the next subsection, we apply these strategies for illustration.

\section{Case studies and a real example} \label{SecCase}
\subsection{Maximin designs}
We consider three cases with $(Q, L) = (1, 255)$, $(2, 242)$, and $(3, 255)$.
For each case, $\mathrm{ISI}$ is set to 4 s and $\mathrm{TR}$ is 2 s. Following \citet
{Kaoetal2009ar51} and \citet{Liu2004ar51},
a second-order polynomial drift in the BOLD times series and an AR(1)
noise with an autocorrelation
coefficient $\rho= 0.3$ are assumed. With these settings, we adapt the
knowledge-based genetic algorithm (GA) of
\citet{Kaoetal2009ar51} to search for maximin designs; see
the supplementary document [\citet{Kaoetal2013ar51}] for the details of
this GA. During the search,
the minima of $\Phi_A$ of candidate designs are evaluated
over a grid on the specific parameter space. The grid interval is 0.2
for $\pb$, and
0.1$\pi$ for $\varphi_i$'s (and thus $\thetab$). When comparing the
obtained designs, finer
grid intervals of 0.1 and 0.05$\pi$ are considered for $\pb$ and
$\thetab$, respectively. We
implement our MATLAB programs on a desktop computer with a 3.4 GHz Core
i7-2600 processor.
These programs are available from the authors.

For $Q = 1$, our GA search first targets a design maximizing $\min_{\mathcal{P}} \Phi_A(d; 1, \pb)$.
Although we focus only on $\{1\} \times\mathcal{P}$, Lemmas \ref{lemma1} and \ref{lemma2}
warrant that
the obtained maximin designs are for the entire parameter space
$\mathbb{R}^1 \times\mathcal{P}$.
For $Q = 2$ and $3$, we follow the proposed strategies to first
identify a subset
$\Theta_0$ of $\Theta$ and a class of permutation matrices $\mathcal
{G}$. We recommend to include
all the $Q$-by-$Q$ permutation matrices, except for the identity
matrix, in $\mathcal{G}$ to allow
a small $\Theta_0$. We take $\Theta_0$ as
$\{(\cos\varphi_1, \sin\varphi_1) \mid\varphi_1 \in[-\pi/4, \pi
/4]\}$ for $Q = 2$,
and as $\{(\cos\varphi_1, \pm\sin\varphi_1 \cos\varphi_2, \pm
\sin\varphi_1 \sin\varphi_2)
\mid\varphi_1 \in[0, \arccos(1/\sqrt{3})]$, $\varphi_2 \in
[\kappa, \pi/4]\}$ for $Q = 3$;
$\kappa= \arccos(\cos\varphi_1/\sin\varphi_1)$ if $\varphi_1 >
\pi/4$, and $\kappa= 0$, otherwise.
We note that, for these two cases, the $\Theta$ defined after Lemma \ref{lemma2}
can be written as $\Theta= \bigcup_{g = 0}^{Q!-1} \Theta^*_g$, where
$\Theta^*_g = \{\tau_{g,\theta} \Gb_g \thetab\mid\thetab\in
\Theta_0\}$, and $\tau_{g,\theta}$ is the sign of
$((\Gb_g \thetab))_1$, the first element of $\Gb_g \thetab$; we set
$\tau_{g,\theta}$ to 1 when $((\Gb_g \thetab))_1 = 0$.
It is easy to see that, by using Lemma \ref{lemma2}, Lemma \ref{lemma3} still holds after replacing
$\Theta_g$ with $\Theta^*_g$.

With the selected $\Theta_0$ for $Q > 1$, the GA is applied to search
for $d_{\Mm, \Theta_0}$ maximizing
$\min_{\Theta_0\times\mathcal{P}} \Phi_A(d; \thetab, \pb)$.
Both Strategies \ref{strategy1} and \ref{strategy2} are considered to reduce computational burden.
Specifically, following Strategy \ref{strategy1}, we apply the GA to find $d_{\Mm,
\Theta_0}$ over the space $\Xi$
of all designs, and obtain the minimal $\mathcal{R}_g$ in (\ref{Eqg0ratio})
as a lower bound of the efficiency of the obtained design. We also use
the GA to find such
a design over the restricted design class $\Xi_0$ (i.e., Strategy \ref{strategy2}).
For each case, we generate ten
designs by using different random seeds in the GA.

\begin{table}
\caption{The performance and mean CPU time (in minutes) for $d_{\Mm}$
obtained over $\Xi$ (all designs) for $Q=1$;
and $d_{\Mm, \Theta_0}$ obtained over $\Xi$ and the subclass $\Xi_0$
for $Q = 2$ and $3$} \label{5Tab1}
\begin{tabular*}{\tablewidth}{@{\extracolsep{\fill}}l d{2.2}d{2.2}d{2.2}d{3.2}d{2.2}@{}}
\hline
& \multicolumn{1}{c}{$\bolds{Q = 1}$} & \multicolumn{2}{c}{$\bolds{Q=2}$} &
\multicolumn{2}{c}{$\bolds{Q=3}$}\\[-4pt]
& \multicolumn{1}{c}{\hrulefill} & \multicolumn{2}{c}{\hrulefill}
& \multicolumn{2}{c@{}}{\hrulefill}\\
& \multicolumn{1}{c}{$\bolds{\Xi}$} & \multicolumn{1}{c}{$\bolds{\Xi}$}
& \multicolumn{1}{c}{$\bolds{\Xi_0}$} & \multicolumn{1}{c}{$\bolds{\Xi}$} &\multicolumn{1}{c@{}}{$\bolds{\Xi_0}$}\\
\hline
$\min$-$\Phi_A$ \\
\quad Maximum & 75.67 & 25.67 & 25.64 & 14.19 & 14.41 \\
\quad Mean & 75.34 & 25.48 & 25.54 & 14.06 & 14.29 \\
\quad Std. err. & 0.08 & 0.05 & 0.04 & 0.03 & 0.03 \\
[3pt]
Mean CPU time & 0.85 & 9.29 & 4.16 &165.96 & 56.54 \\
\hline
\end{tabular*}
\end{table}

Table~\ref{5Tab1} presents the maximum, mean, and standard error of
$\min$-$\Phi_A$ of the ten GA-generated designs.
The mean CPU time for obtaining these designs is also reported. As in
Table~\ref{5Tab1}, our two strategies
yield designs with similar $\min$-$\Phi_A$ values. In addition, the
minimal $\mathcal{R}_g$ for the
designs obtained with Strategy \ref{strategy1} is at least $98.78\%$ for $Q=2$ and at
least $97.99\%$ for $Q=3$,
indicating that our obtained designs are very efficient compared
with a maximin design $d_{\Mm}$ for $\Theta\times\mathcal{P}$. We
also note that a direct search
for $d_{\Mm}$ can be very time consuming for $Q>1$. By focusing on the
reduced parameter space $\Theta_0 \times\mathcal{P}$,
our proposed methods can efficiently generate high quality designs. In
addition, obtaining maximin designs over the
subclass $\Xi_0$ of designs can further reduce the computational
burden without having a negative effect on the design efficiencies.
These results provide compelling evidence for the efficiency and
effectiveness of Strategy \ref{strategy2}.

\begin{figure}

\includegraphics{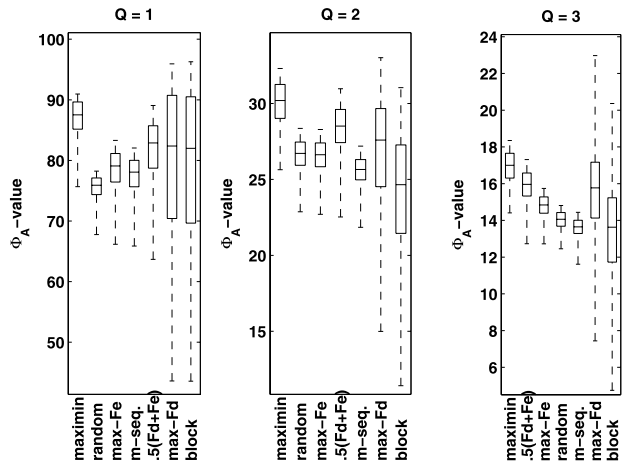}

\caption{Boxplots of the $\Phi_A$-values over $\Theta\times\mathcal{P}$
of the competing designs, including maximin designs,
designs obtained with linear models for estimation ($\max$-$F_e$),
detection ($\max$-$F_d$) and both ($0.5(F_d+F_e$)), a block design, an
$m$-sequence,
and the design maximizing $\min$-$\Phi_A$ selected from 100 randomly
generated designs. Designs
are ordered by their $\min$-$\Phi_A$ values.} \label{Fig0}
\end{figure}

We also compare the obtained designs with some traditional designs that
are widely used in practice.
Figure~\ref{Fig0} presents the boxplots of the average estimation
precision ($\Phi_A$)
over $\Theta\times\mathcal{P}$ for the competing designs. In that
figure, the selected maximin design for $Q = 1$
is the design maximizing $\min$-$\Phi_A$ over the ten $d_{\Mm}$
designs obtained by the GA;
for $Q > 1$, the selected designs maximize $\min$-$\Phi_A$ over the
ten $d_{\Mm, \Theta_0}$ designs
obtained via Strategy~\ref{strategy2}. The traditional designs include block designs,
$m$-sequences,
$\max$-$F_d$, $\max$-$F_e$, bi-objective, and random designs. Block
designs for fMRI are sequences
formed by repetitions of $\{B_0 B_1 B_2\cdots B_Q\}$, where $B_q$ is a
sequence of $q$'s (i.e., $\{qq\cdots q\}$)
of a given size. Here, we consider block designs of size four that are
formed by $\{00001111\cdots$ $QQQQ\}$. Under linear models, these
designs can yield high performance
for detecting brain activation [Maus et~al.
(\citeyear{Mausetal2011ar51,Mausetal2010ar52}), \citet{Henson2007bc51}].
An $m$-sequence can be generated from primitive
polynomials for a Galois field [\citet
{BuracasBoynton2002ar51,Godfrey1993bk51,MacWilliamsSloane1977bk51}].
These designs can be obtained from
a MATLAB program provided by \citet{Liu2004ar51} and are good for
estimating the HRF.
The $\max$-$F_d$, $\max$-$F_e$, and bi-objective ($0.5F_d +0.5F_e$)
designs are obtained by the
GA of \citet{Kaoetal2009ar51} with linear models. A $\max$-$F_d$
design maximizes the efficiency of detection,
whereas a $\max$-$F_e$ design maximizes the HRF estimation efficiency.
The bi-objective designs maximize the
average of theses two efficiencies. They offer a compromise between the
two competing objectives of
detection and estimation. We also generate 100 random designs and
select the one
yielding the maximal $\min$-$\Phi_A$. When lacking design tools for
sophisticated
experimental settings, as considered here, random designs are not
uncommon in practice.
More details about these designs can be found in \citet
{Kaoetal2009ar51} and \citet{Liu2004ar51}.

The designs in Figure~\ref{Fig0} are ordered by their $\min$-$\Phi
_A$ values.
Clearly, the \mbox{maximin} designs are much better than the other designs and
have relatively small dispersions in $\Phi_A$-values across the
parameter space. This indicates that the estimation precisions yielded
by the maximin designs
are quite robust against a misspecified parameter vector value. We also
observe that, while the block and $\max$-$F_d$ designs are
recommended for detecting activation under linear models, they do not
perform well
for detection under the nonlinear model. The $\Phi_A$-value of these
two types of designs
can vary greatly over the parameter space, and, at the worst cases,
their $\Phi_A$-values
can be very low, indicating poor precisions in estimating $\thetab$.

\subsection{Maximin-efficient designs}
Our proposed methods are also applied to obtain maximin-efficient
designs. For $Q = 1$,
we first use the GA to search for the required locally optimal designs
for each grid point on $\{0,1\} \times\mathcal{P}$,
and then a design maximizing $\min_{\{0,1\} \times\mathcal{P}} \operatorname{RE}(d;
\thetab, \pb)$.
Based on Corollary~\ref{coroll2}, the GA actually yields a maximin-efficient design
$d_{\mathrm{MmE}}$ for $\mathbb{R}^1 \times\mathcal{P}$
even though the reduced parameter space is considered. For $Q = 2$ and
$3$, we consider the $\Theta_0$ presented in
the previous subsection. We then apply the GA to search for (1) locally
optimal designs over
$\{\{\zerob\} \cup\Theta_0\} \times\mathcal{P}$; and (2) a
maximin-efficient designs $d_{\mathrm{MmE}, \Theta_0}$
optimizing $\min_{\{\{\zerob\} \cup\Theta_0\} \times\mathcal
{P}} \operatorname{RE}(d; \thetab, \pb)$. The $d_{\mathrm{MmE}, \Theta_0}$ designs are
obtained over the entire design space $\Xi$ (Strategy~\ref{strategy1}) and over the
subclass $\Xi_0$ (Strategy \ref{strategy2}).

\begin{table}
\caption{The performance and mean CPU time (in minutes) for $d_{\mathrm{MmE}}$
obtained over $\Xi$ (all designs) for $Q=1$;
and $d_{\mathrm{MmE}, \Theta_0}$ obtained over $\Xi$ and the subclass $\Xi_0$
for $Q = 2$ and $3$} \label{5Tab11}
\begin{tabular*}{\tablewidth}{@{\extracolsep{\fill}}l ccccc@{}}
\hline
& \multicolumn{1}{c}{$\bolds{Q = 1}$} & \multicolumn{2}{c}{$\bolds{Q=2}$} &
\multicolumn{2}{c}{$\bolds{Q=3}$}\\[-4pt]
& \multicolumn{1}{c}{\hrulefill} & \multicolumn{2}{c}{\hrulefill}
& \multicolumn{2}{c@{}}{\hrulefill}\\
& \multicolumn{1}{c}{$\bolds{\Xi}$} & \multicolumn{1}{c}{$\bolds{\Xi}$}
& \multicolumn{1}{c}{$\bolds{\Xi_0}$} & \multicolumn{1}{c}{$\bolds{\Xi}$} &\multicolumn{1}{c@{}}{$\bolds{\Xi_0}$}\\
\hline
$\min$-$\mathrm{RE}$ \\
\quad Maximum & 0.835 & \hphantom{0}0.790 & 0.829 & \hphantom{00}0.797 & \hphantom{0}0.829 \\
\quad Mean & 0.830 & \hphantom{0}0.783 & 0.820 & \hphantom{00}0.783 & \hphantom{0}0.823 \\
\quad Std. err. & 0.001 & \hphantom{0}0.002 & 0.002 & \hphantom{00}0.003 & \hphantom{0}0.001 \\
[3pt]
Mean CPU time & 0.88\tabnoteref{ta} & 11.69\tabnoteref{tb} &
5.88\tabnoteref{tb} & 207.51\tabnoteref{tc} & 52.04\tabnoteref{tc} \\
\hline
\end{tabular*}
\tabnotetext[\mbox{1}]{ta}{Additional 17 minutes are needed for finding the
required locally optimal designs.}
\tabnotetext[\mbox{2}]{tb}{Additional 4 hours are needed for finding the
required locally optimal designs.}
\tabnotetext[\mbox{3}]{tc}{Additional 46 hours are needed for finding the
required locally optimal designs.}
\end{table}

Table~\ref{5Tab11}, to be read as Table~\ref{5Tab1}, presents a
comparison among the obtained maximin-efficient designs.
By omitting the time needed for obtaining locally optimal designs, the
CPU times
in Table~\ref{5Tab11} for obtaining the maximin-efficient
designs are similar to those for maximin designs in Table~\ref{5Tab1}. However,
obtaining maximin-efficient designs requires locally optimal designs.
This unfortunately
makes maximin-efficient designs computationally much more expensive
than maximin designs,
especially when $Q$ becomes large. Specifically, for $Q = 1$, we use
the GA to obtain 352 locally optimal
designs, each requiring about 2.88 s. The GA takes about 4 hours to
find 1232 locally optimal
designs for $Q = 2$, and about 46 hours to generate 5984 locally
optimal designs for $Q = 3$. Results in Table~\ref{5Tab11} also
indicate that
Strategy \ref{strategy2} outperforms Strategy \ref{strategy1} in terms of the achieved design efficiency
and required CPU time. Strategy \ref{strategy2} is thus recommended.

\begin{figure}

\includegraphics{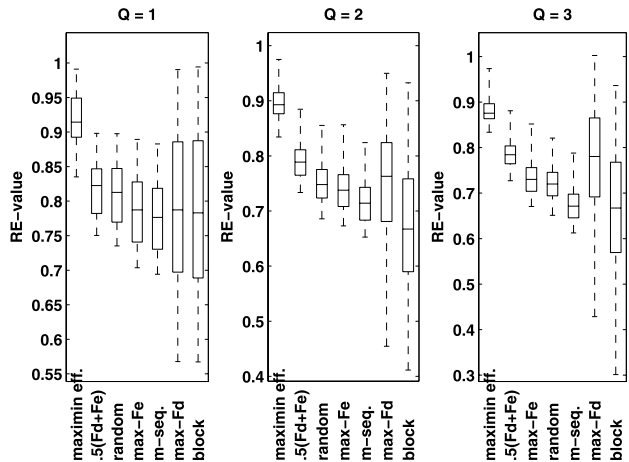}

\caption{Boxplots of the $\mathrm{RE}$-values over $\Theta\times\mathcal{P}$
of maximin-efficient designs, $\max$-$F_e$, $\max$-$F_d$, and
bi-objective ($0.5(F_d+F_e$)) designs, a block design, an $m$-sequence,
and the design maximizing $\min$-$\mathrm{RE}$ selected from 100 randomly
generated designs. Designs
are ordered by their $\min$-$\mathrm{RE}$ values.} \label{Fig01}
\end{figure}

In Figure~\ref{Fig01} we compare the $\mathrm{RE}$-values over $\Theta\times
\mathcal{P}$ of
the maximin-efficient designs and the traditional designs introduced in the
previous subsection. The selected maximin-efficient design for $Q = 1$ maximizes
$\min$-$\mathrm{RE}$ over the ten $d_{\mathrm{MmE}}$ designs; the maximin-efficient
designs for $Q > 1$ are selected from the ten $d_{\mathrm{MmE}, \Theta_0}$ designs
obtained by Strategy \ref{strategy2}. As presented in the figure, our designs
significantly outperform the traditional designs.

\subsection{An example} \label{SecExample}
In this subsection we consider an experimental setting employed by
\citet{Miezinetal2000ar51}, in which
a 1.5-s 8-Hz flickering checkerboard (stimulus) is presented interlaced
with a visual fixation (control).
Upon the onset of each checkerboard, subjects responded by pressing a
key with their right hands.
The minimal time between consecutive stimulus onsets was 2.5 s ($\mathrm{ISI} =
2.5$ s). The BOLD time series was acquired
every 2.5 s ($\mathrm{TR} = 2.5$ s). The experimenters presented the same design
twice to a subject with a 2-minute rest period
in between the two runs. Each run lasted about 5.5 minutes. To allow an
effective sampling rate of the hemodynamic response,
stimulus onsets were synchronized with MR scans in the first run and
were shifted 1.25 s in the second run.

Miezin and colleagues demonstrated that the time-to-peak ($p_1$) and
time-to-onset ($p_6$) of the HRF
can vary across brain voxels. Taking this uncertainty into account, we
apply our proposed approach
to obtain maximin and maximin-efficient designs. A simple modification
is needed
to accommodate the special requirement that the study is conducted over
two runs.
Specifically, we replace the design matrix in model (\ref{EqNLM}) by
$\operatorname{diag}(\Xb_{d,1}, \Xb_{d,1})$ since the same sequence of stimuli is
presented twice. In addition,
$\hb(\pb)$ is now $(\hb_1(\pb)', \hb_2(\pb)')'$, where the $j$th
element of $\hb_1(\pb)$ is
$g(2.5(j-1); \pb)$ and that of $\hb_2(\pb)$ is $g(1.25+2.5(j-1); \pb
)$. This
accounts for the difference of 1.25 s in the HRF sampling time points
between the two runs. We also
consider the nuisance term $[(\Sb\gammab_1)', (\Sb\gammab_2)']'$ that
allows run effects, where $\Sb\gammab_i$ corresponds to a
second-order polynomial drift, $i = 1, 2$.
The noise of the two runs are assumed to be two independent AR(1)
processes with autocorrelation coefficient
$\rho= 0.3$. The whitening matrix thus has the form $(\Ib_2 \otimes
\Vb)$, where $\Vb$ is a whitening matrix
for each run [see also \citet{Kaoetal2009ar52}]. We also investigated
the performance of
our obtained designs when $\rho=0$ or $0.5$, and found that our
designs are still quite efficient
with a different $\rho$-value.

\begin{figure}

\includegraphics{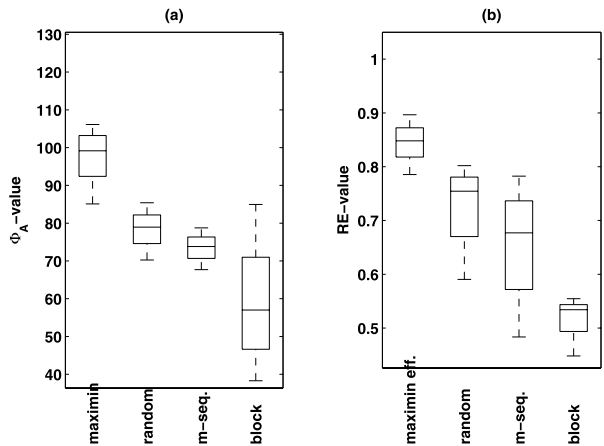}

\caption{Boxplots of the \textup{(a)} $\Phi_A$-values, and \textup{(b)} $\mathrm{RE}$-values of
the competing designs over $\{1\}\times\mathcal{P}$.}
\label{FigExpBoxP}
\end{figure}

In addition to maximin-type designs, we generate a block design,
$m$-sequence-based design,
and 100 designs that are random permutations of a design consisting of
50\% ``0''s and 50\% ``1''s.
The block design is formed by repeating \{000000111111\}, which
has the 15s-off-15s-on pattern that is often recommended for detecting
brain activation. An $m$-sequence does not exist in this case.
We thus follow \citet{Liu2004ar51} to generate an $m$-sequence-based
design by concatenating an $m$-sequence
of length 127 with its first 5 elements. The 100 randomly permuted
designs are constructed to mimic
the design considered by \citet{Miezinetal2000ar51}. Each of these 100
designs are selected so that the average time
between consecutive stimulus onsets is within 4.9 s and 5.1 s. Among
these 100 designs, we select the one
yielding the maximal $\min$-$\Phi_A$ value when comparing with the
maximin design, and the one maximizing
$\min$-$\mathrm{RE}$ when comparing with the maximin-efficient design. Figure~\ref{FigExpBoxP} provides summaries
of the performances of these designs. As shown in the figure, our
proposed methods consistently generate high-quality designs that
significantly outperform the traditional designs.

We also explore the performance of the designs with different $\rho$-values.
When $\rho= 0$, the maximin design for $\rho=0.3$ attains $95.1\%$ of
the $\min$-$\Phi_A$ value
of the maximin designs for $\rho= 0.0$.
The $\min$-$\mathrm{RE}$ value of maximin-efficient design for $\rho=0.3$ is
$98.1\%$ of
that of the maximin-efficient design for $\rho=0.0$. For cases where
$\rho$ is as
large as $\rho= 0.5$, the relative $\min$-$\Phi_A$ of the maximin
design for $\rho= 0.3$
to the maximin design for $\rho= 0.5$ is $97.2\%$. For this same
condition, the maximin-efficient
design for $\rho= 0.3$ retains $92.0\%$ of the $\min$-$\mathrm{RE}$ of the
maximin-efficient
design for $\rho= 0.5$. Our obtained designs, especially the maximin designs,
perform relatively well when comparing with the best design for a $\rho$-value
that is as small as $0$ and as high as $0.5$.

\section{Conclusions}\label{sec4}
We obtain high-quality experimental designs for fMRI experiments
to help to render efficient statistical inference on brain activity
with a nonlinear model. In contrast to linear models, the nonlinear model
allows us to detect brain voxels activated by the mental stimuli while the
uncertain HRF shape is taken into account. However, optimal designs
for the nonlinear model depend on unknown model parameters, making the design
problem notoriously difficult. To tackle this problem, we consider
maximin and maximin-efficient designs and propose efficient approaches
for obtaining these designs. Our approaches involve a large reduction
in the parameter space, a
restricted class of ER-fMRI designs, and the use of the knowledge-based
GA of
\citet{Kaoetal2009ar51} for searching for maximin-type
designs. These approaches, especially Strategy \ref{strategy2}, are demonstrated to
be powerful via case studies
and a real example.

Maximin and maximin-efficient designs are widely accepted, although
obtaining them is almost always difficult.
\citet{PronzatoWalter1988ar2071}
studied both types of designs and concluded that maximin designs have
definite advantages over maximin-efficient designs
when reducing the worst possible uncertainty for estimating the
parameter is of concern.
On the other hand, \citet{DetteBiedermann2003ar2071} considered the
maximin-efficient criterion
because it tends to avoid placing too much attention on a certain
parameter vector value.
\citet{HuangLin2006ar2071} suggested that both maximin-type criteria
deserve consideration;
the selection may thus be guided by the need and preference of the
experimenter. Our approaches allow us to efficiently find both maximin
and maximin-efficient designs,
although the latter designs are computationally more expensive than the
former designs.

We also observe that, while blocked designs are recommended for
detecting brain activation under
linear models, they are very poor for the same objective under
nonlinear models with uncertain HRF shapes.
We believe that the inferiority of block designs is mainly due to
their low efficiencies in estimating the HRF [e.g., \citet{LiuFrank2004ar51}].
The estimation of the HRF shape, although not the main concern, is needed
when the HRF shape is uncertain. A good design should thus allow
for a reasonable efficiency in performing this task. Designs with
random components tend to serve this purpose well.

The designs that we found are for a nonlinear model, in which the HRF
is approximated by
$\theta\hb(\pb)$, the product of the unknown HRF amplitude $\theta$
and the uncertain HRF shape $\hb(\pb)$.
Such models are not uncommon in the literature [\citet
{LindquistWager2007ar51,Handwerkeretal2004ar51,Miezinetal2000ar51}].
While we focus on
an $\hb(\pb)$ having the same form as the popular double-gamma
function of SPM, the proposed
approaches can be extended to $\hb(\pb)$'s of other forms, for example,
the inverse logit function considered by \citet{LindquistWager2007ar51}.

When implementing our approaches in the case studies, we consider an
AR(1) noise with known constant
autocorrelation coefficient $\rho=0.3$. This selection is guided
mainly by previous studies, for example, \citet{Lenoskietal2008ar51}
and \citet{Worsleyetal2002ar51}. From these studies, the use of AR(1)
noise tends to provide
satisfactory analysis results. With AR(1), the results of \citet
{Mausetal2010ar51}
for linear models suggest that the obtained designs for $\rho=0.3$ do not
suffer a significant loss in design efficiency under other values of
$\rho\in[0, 0.5]$.
We also observe a similar outcome under the nonlinear model.

We also note that the assumed AR(1) model with $\rho=0.3$
could be idealistic. First, other models for
autocorrelated noise might be more appropriate for some data [e.g.,
\citet{Lindquist2008ar51}].
In addition, whether for an AR(1) model or another model,
knowledge about the unknown, possibly nonconstant parameter(s)
may not always be available at the design stage.
For a selected model, our methods could then be
combined with the approach of \citet{Mausetal2010ar51}
to search for optimal designs (using the maximin or maximin-efficient criterion)
by taking uncertain HRF shape and autocorrelation parameters into account.
While such a maximin-type approach can be generalized to accommodate
different models for autocorrelated noise, the design problem can
become very challenging. Developing an efficient method for cases where
both HRF shape and correlation are uncertain is a topic of future research.

\begin{appendix}\label{app}
\section*{\texorpdfstring{Appendix: A proof of Lemma \lowercase{\protect\ref{lemma3}}}
{Appendix: A proof of Lemma 3}}
The following two lemmas are straightforward.
Their proofs are thus omitted. The notation used is as in Lemma \ref{lemma3}.

\begin{lemma}\label{lemmaA.1}
For a permutation matrix $\Gb_g$, let
$k_{G_g}(d)$ be the design
obtained by relabeling the stimulus types, the same way as $\Gb_g$
permutes $(1,2,\ldots,\break  Q)'$, of a design $d$. We have $\Mb(k_{G_g}(d); \Gb
_g\thetab, \pb) = \Gb_g'\Mb(d; \thetab, \pb)\Gb_g$ and, thus,
$\Phi_A(k_{G_g}(d);\allowbreak \Gb_g\thetab, \pb) = \Phi_A(d; \thetab, \pb
)$ for any $(\theta, \pb) \in\Theta_0 \times\mathcal{P}$.
\end{lemma}

\begin{lemma}\label{lemmaA.2}
The following two conditions are equivalent: (1)
$d_0^*$ is a maximin design for $\Theta_0 \times\mathcal{P}$;
and (2) $k_{G_g}(d_0^*)$ is a maximin design for $\Theta_g \times
\mathcal{P}$ for any $g$.
\end{lemma}

\begin{pf*}{Proof of Lemma \ref{lemma3}}
For a $d_{\Mm, \Theta_0}$ satisfying the conditions of
Lemma~\ref{lemma3}, we have $\min_{\Theta_g
\times\mathcal{P}}\Phi_A(k_{G_g}(d_{\Mm, \Theta _0}); \thetab, \pb) =
\min_{\Theta_0 \times\mathcal{P}} \Phi_A(d_{\Mm, \Theta_0}; \thetab,
\pb) =\break  \min_{\Theta_g \times\mathcal{P}} \Phi_A(d_{\Mm, \Theta_0};
\thetab, \pb)$ for any $g$. Therefore, $d_{\Mm, \Theta_0}$, which is a
maximin design for $\Theta_0 \times\mathcal{P}$, is also a maximin
design for $\Theta_g \times\mathcal{P}$ for any $g$, and for $
(\bigcup_{g=0}^{G}\Theta_g ) \times\mathcal{P} =
\Theta\times\mathcal{P}$.
\end{pf*}
\end{appendix}

\section*{Acknowledgments}
We thank anonymous referees for raising questions that resulted
in an improvement of this article.

\begin{supplement}
\stitle{Supplement to ``Maximin and maximin-efficient event-related FMRI designs under
a nonlinear model''}
\slink[doi]{10.1214/13-AOAS658SUPP} 
\sdatatype{.pdf}
\sfilename{aoas658\_supp.pdf}
\sdescription{We provide (1) a proof that the $\Phi_A$-value of a
design in the restricted design class $\Xi_0$ is
insensitive to permutations of the elements of $\thetab$; and (2) a genetic
algorithm for obtaining ER-fMRI designs.}
\end{supplement}


\printaddresses

\end{document}